\documentstyle[graphicx,twocolumn,eqsecnum,aps]{revtex}

\begin{document}
\title{Microelectromagnets for Trapping and Manipulating \\Ultracold Atomic Quantum Gases
}
\author{J.\ Fort{\'a}gh, H.\ Ott, G.\ Schlotterbeck, and C.\ Zimmermann}
\address{
Physikalisches Institut der Universit{\"a}t T{\"u}bingen\\
Auf der Morgenstelle 14, T{\"u}bingen, Germany}
\author{B.\ Herzog, and D.\ Wharam}
\address{
Institut f{\"u}r Angewandte Physik, Universit{\"a}t T{\"u}bingen\\
Auf der Morgenstelle 10, T{\"u}bingen, Germany}
\date{\today}
\maketitle
\begin{abstract}
We describe the production and characterization of microelectromagnets made 
for trapping and manipulating atomic ensembles. The devices consist of 
7 fabricated parallel copper conductors 3\,$\mu$m thick, 25\,mm long, with 
widths ranging from 3\,$\mu$m to 30\,$\mu$m, and are produced by 
electroplating a sapphire substrate. Maximum current densities in the wires 
up to $6.5\times10^6$~A\,cm$^{-2}$ are achieved in continuous mode operation. The 
device operates successfully at a base pressure of 10$^{-11}$\,mbar. The 
microstructures permit the realization of a variety of magnetic field 
configurations, and hence provide enormous flexibility for controlling the motion 
and the shape of Bose-Einstein condensates. 
\end{abstract}

\narrowtext

The combination of ultracold atoms and miniaturized magnetic traps \cite{first}   
suggests fascinating scenarios for integrated atom optic devices including waveguides, 
interferometers, resonators, and quantum gates \cite{second}. 
The current progress in reducing the 
production time of Bose-Einstein condensates below 1\,s \cite{third} and the recently achieved 
significant simplification of the magnetic apparatus \cite{fourth} adds to the vision of a future 
quantum technology with coherent matter waves analogous to today's highly developed 
laser technology. One of the crucial prerequisites is the preparation of atomic clouds 
in steep potentials. As a particularly attractive approach, magnetic fields 
generated by miniaturized current conductors can be used \cite{fifth}.  
In addition, the 
trapping potentials can be spatially structured on a micron length scale and temporal 
modulations are possible on the scale of microseconds. By using miniaturized 
conductors, 
the trapping potential is located close to the surface of the substrate. 
At a distance of about 
100\,$\mu$m from the surface loss mechanisms, heating and decoherence 
effects within the cloud are 
predicted to become significant \cite{sixth}.  
Earlier experiments succeeded in loading atoms into microtrap 
potentials that were generated by current conductors on the scale of 
several 10\,$\mu$m \cite{seventh}. 
Microtraps based on smaller conductors have been demonstrated electrically \cite{eighth},  
and single atoms have been successfully guided in a micro magnetic channel \cite{ninth}.  
In this Letter we describe the realization of 
magnetic microtraps with ultra thin conductors that reach the limit of 
surface-atom-interaction. Details of the fabrication procedure, and the chemical and 
electrical properties are given. The particular microstructure described has been used for 
generating highly anisotropic Bose-Einstein condensates \cite{fourth}. We give a brief 
summary of the experiments and conclude with an estimation of the field configurations which 
can be reached with this device.

The microelectromagnet consists of a parallel configuration of 
copper wires with widths ranging from 3\,$\mu$m to 30\,$\mu$m.  The thickness of the 
galvanically 
deposited copper was of the order of a few microns, while the length of the 
total structure was approximately 25 mm.  
\begin{figure} \centering
\includegraphics[width=6.8cm]{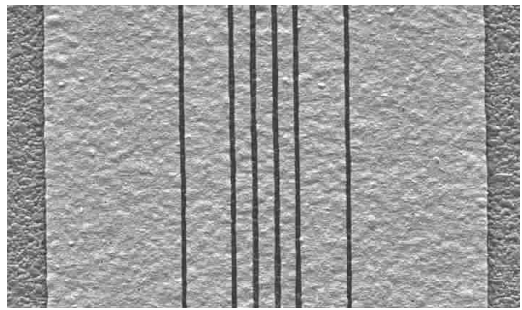}\\
\vspace*{4mm}
\caption{An electron micrograph of a typical microtrap showing a segment of 
the wire geometry employed. 
The wire widths are 30\,$\mu$m, 11\,$\mu$m, and 3\,$\mu$m, and the nominal separation between 
wires is 1\,$\mu$m.
The quality of the electroplating can be seen in the smoothness of the copper wires. 
The residual roughness is presumably due to the roughness of the sapphire substrate. 
\label{Fig1}}\end{figure}
The fabrication of thin metallic wires with 
such high aspect ratios and excellent conducting properties requires the use of sophisticated 
microfabrication techniques.  Initially, a high-quality optical mask was prepared using 
electron-beam lithography, which was then employed to transfer the wire pattern onto the 
smooth (residual roughness $\le 1\,\mu$m) sapphire substrate.  
In order to guarantee good adhesion 
and to provide a contact layer for the subsequent galvanic processing the substrate was also 
coated with a thin (7\,nm Cr followed by 120\,nm Cu) metallic layer by thermal evaporation.  
The wire fabrication itself was performed in a conventional acidic CuSO$_4$ galvanic bath 
with an additional organic brightener which produces copper layers of 
high-purity ($\ge 99.99\,\%$) 
and a density of 8.9\,g\,cm$^{-3}$.  Best results were obtained 
using the technique of pulsed plating where the current was modulated with a square 
wave of frequency 1\,kHz and current values of 360\,mA and -40\,mA for the positive and 
negavtive excursions respectively.  This technique is 
well suited for the fabrication of smooth metallic layers as evidenced by the finished 
structure shown in Fig.\,\ref{Fig1}.

For the chosen cathode area (ca.\,44\,cm$^2$) the above conditions correspond to a current density 
of approximately 0.8\,A\,dm$^{-2}$ (and -0.09\,A\,dm$^{-2}$ for the negative half-cycle), 
and to an 
electroplating rate of 1\,nm\,s$^{-1}$.  After removal of the photoresist the metallic contact 
layer was chemically etched with nitric acid for copper removal followed by potassium 
ferricyanide for the chrome removal.  The quality of the electroplated layer and substrate 
was 
confirmed via EDX measurements which show neither the presence of magnetic impurities nor 
significant contamination.

\begin{figure} \centering
\includegraphics[width=6.8cm]{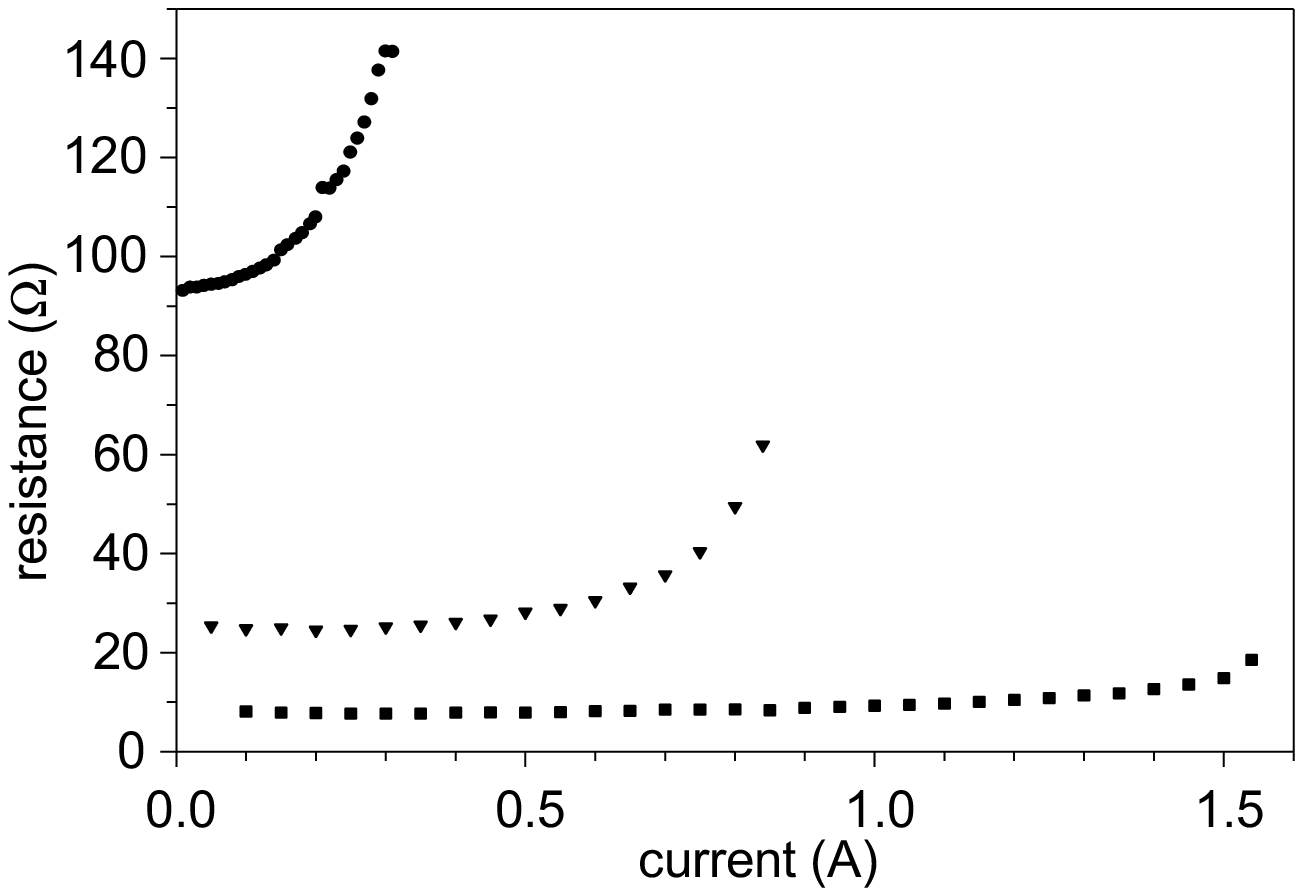}\\
\vspace*{4mm}
\caption{The device resistance is plotted as a function of the driving current for three 
different wire widths 3\,$\mu$m($\circ$), 11\,$\mu$m ($\bigtriangledown$), 
and 30\,$\mu$m ($\Box$). The 
charcteristics stop at the last measureable point before thermal damage leads to device 
breakdown.  
\label{Fig2}}\end{figure}

A test structure with a 1.6\,$\mu$m copper layer was used to determine the critical 
current at which the conductors are thermally damaged. The measurements were carried out at 
room temperature in air with the microstructure mounted on a copper heat sink. The electrical 
contact to the microstructure is accomplished by bonding 3 aluminium wires with 20\,$\mu$m 
diameter and a length of about 700\,$\mu$m onto each contact. 
Roughly 10\,s after changing the current in the conductors a steady state 
temperature is reached, and the voltage drop along the wires can be recorded. 
Thus the current 
was increased every 10\,s in steps of 10\,mA  for  the 3\,$\mu$m-conductor and in steps of 
50 mA for 11\,$\mu$m- and 30\,$\mu$m-conductors. The maximum stable current is marked by the 
last data point in Fig.\,\ref{Fig2}, which shows the device resistance as a function of 
the current. 
The corresponding maximum current densities are  
3.1$\times10^6$\,A\,cm$^{-2}$, 
4.8$\times10^6$\,A\,cm$^{-2}$, and 
6.5$\times10^6$\,A\,cm$^{-2}$ for the 30\,$\mu$m, 11\,$\mu$m, and 3\,$\mu$m conductors respectively. 
The maximum current density increases with decreasing width, presumably because the geometry 
of thinner conductors favours a more efficient heat contact with the substrate.    
The resistance 
increases by a factor of 2.5 and of 2.3 respectively which corresponds to a 
temperature increase of about 350\,K, determined using the thermal coefficient of resistance for 
bulk copper ($\alpha = 0.004\,$K$^{-1}$). We therefore estimate that the average temperature of 
the wires at breakdown is roughly 650\,K. Near imperfections, which may be caused by 
unintentional variations of the conductor cross section or by reduced adhesion to the 
substrate, we expect the temperature to 
exceed the 
average value and to lead to breakdown when the local temperature reaches 
the melting point of bulk copper (1350\,K). 
The temperature behaviour near the maximum current is similar for the 
11\,$\mu$m and 30\,$\mu$m conductors. The resistance 
of the 3\,$\mu$m conductor increases only by a factor of 1.5 at the maximum current and the 
average temperature is increased to only 450\,K.  This can be explained by a stronger 
influence of the imperfections. The maximum current 
densities measured in our test experiments at room temperature are similar to the critical 
current densities in high-T$_{\rm c}$ superconductors at liquid nitrogen temperature 
\cite{fifth} and fall short only by a factor 15 of the values obtained by 
Drndic et al.\,\cite{eighth} with gold conductors 
on sapphire substrates at liquid helium temperature. The microelectromagnets are 
used under ultra-high vacuum conditions and
can be cooled with liquid nitrogen. A significant resistance reduction and consequent 
increase in critical current density is observed. In addition, the cooling 
reduces outgassing of the surface and results in longer lifetimes of the 
trapped atoms. Nevertheless the longest lifetimes observed in recent experiments 
\cite{twelfth} depend critically upon the separation between substrate and condensate; for 
separations of 200\,$\mu$m and 20\,\,$\mu$m lifetimes of 15\,s and 700\,ms respectively were 
found.

The configuration of seven parallel conductors with different widths combined with the 
extreme length of the microstucture of 25\,mm permits the definition of a variety of magnetic 
trapping potentials \cite{fourth}. 
Besides the usual magnetic traps used for Bose-Einstein condensation,  
very elongated waveguides can also be obtained by changing the current in the microstructure. 
Furthermore, a parallel set of waveguides can be formed and may be merged and separated 
by varying the strength of the offset field perpendicular to the waveguides \cite{thirteen}. 
Such configurations permit the realization of on-chip interferometers, and are currently the 
subject of investigation.  

\begin{figure} \centering
\includegraphics[width=6.8cm]{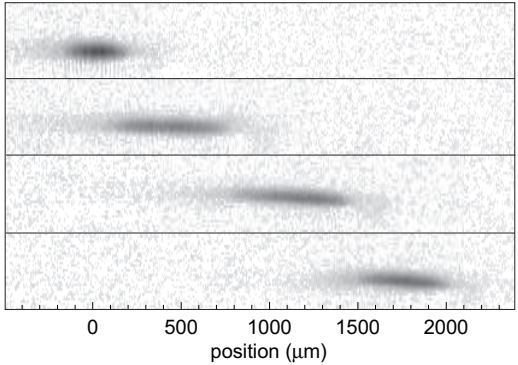}\\
\vspace*{4mm}
\caption{The time evolution of the Bose-Einstein condensate is shown at $t=0\,$ms, $t=100\,$ms, 
$t=200\,$ms, and $t=300\,$ms respectively. The trap was turned off 20\,ms before the first 
image of the condensate was obtained and after the potential modification described in the 
text was complete. The number of atoms in the condensate is estimated to be $1.5\times10^5$.   
\label{Fig3}}\end{figure}

In order to achieve the Bose-Einstein condensation \cite{fourth} one conductor is 
driven for example with an current source of 0.2\,A, and an external bias field of 
4\,G is applied. The resulting gradient in this instance amounts to 400\,G\,cm$^{-1}$ 
and the center of the trap is located 100\,$\mu$m above the surface of 
the microstructure. With a superimposed offset field of 1\,G aligned parallel to the 
conductor the curvature of the magnetic field magnitude in the transverse direction 
amounts to 16\,kG\,cm$^{-2}$ corresponding to a radial (transverse) oscillation 
frequency of $\omega_r = 2\pi\times510$\,s$^{-1}$ for $^{87}$Rb atoms in the 
$|$F=2,m$_{\rm F}$=2$>$ ground state. The axial (longitudinal) confinement of the 
atoms is separately controlled by an additional Ioffe-type electromagnet \cite{tenth}, 
which is adjusted such that the axial oscillation frequency amounts to 
$\omega_a = 2\pi\times14$\,s$^{-1}$. The atoms are loaded into the 
microtrap by adiabatic compression \cite{first} and subsequently cooled below the 
critical temperature by forced rf-evaporation. Condensation is reached with 
10$^6$ atoms at a temperature of 1\,$\mu$K. 
This standard trap configuration can be continuously changed into a much steeper potential 
with an elongated waveguide of large aspect ratio. In the particular experiment shown in 
Fig.\,\ref{Fig3} the axial oscillation frequency of the trap was reduced by a factor of 10 
on a timescale of 400\,ms. During the last 100\,ms the trap center was shifted by  
1\,mm such that the condensate is free to propagate along the waveguide.
This motion is clearly visible, and observations over longer time scales show oscillations of 
the center of mass of the condensate at 
the trap frequency of $2\pi\times1.4$\,s$^{-1}$.
The distance to the surface and the radial oscillation frequency were tuned to 
270\,$\mu$m and $2\pi\times100$\,s$^{-1}$ accordingly. 

In the most extreme potential configuration obtainable with our microstructure, 
defined by the maximum current density and the size of the thinnest 
conductor, the magnetic field gradient is expected to reach 500\,kG\,cm$^{-1}$. 
With a current of 200\,mA in the center 
conductor and -325\,mA in both of the two neighboring conductors the trap is located 3\,$\mu$m 
above the surface and has a depth of 133\,G corresponding to a temperature of 8\,mK for 
$^{87}$Rb 
atoms. With an axial offset field of 1\,G, the radial curvature amounts 
$2.5\times10^{11}$\,G\,cm$^{-2}$ 
corresponding to an oscillation frequency of $\omega_r = 2\pi\times600,000$\,s$^{-1}$. 
By choosing a small axial 
oscillation frequency (for example $\omega_a < 2\pi\times5$\,s$^{-1}$), an aspect ratio of 
$\omega_r/\omega_a >$\,100,000 can be achieved.  
To date, the properties of cold atomic ensembles have not been investigated under such 
extreme conditions due to the lack of an appropriate device. In our most recent experiments 
we observe periodic substructures in the spatial distribution of an elongated condensate 
at a surface separation below 150\,$\mu$m, which is reported in detail 
elsewhere \cite{twelfth,fourteen}. 
The obvious causes for such modulated magnetic or electrostatic potentials have been excluded, 
and we speculate that their origin lies in the magnetic surface properties of the conductors. 

In conclusion, we have demonstrated a device for the manipulation of Bose-Einstein 
condensates based on microfabricated electromagnets. The microfabrication technique employed  
allows the construction of even more sophisticated micromagnets, and opens new possibilities 
for integrated atom-optical devices with ultracold atoms. Novel sensors for rotation and 
for precision measurement of the gavitational field may become feasible. 

\acknowledgments

This work is supported in part by the Deutsche Forschungsgemeinschaft under grant no. 
Zi 419/3-1.  We thank S.\ Raible and R.\ Frank for technical assistance, and 
HighFinesse for providing ultra stable current sources for driving the 
microelectromagnet.

\end{document}